\documentclass[fleqn,twoside]{article}
\usepackage{espcrc2}
\usepackage{cite}
\usepackage{axodraw}
\usepackage{amsmath}
\usepackage{eufrak}

\newcommand{\Li}[2]{{\mbox{Li}}_{#1}\left(#2\right)}
\newcommand{\ep}{\varepsilon}
\newcommand{\veps}{\varepsilon}

\newcommand{\ha}{\frac{1}{2}\:}
\newcommand{\wz}{\sqrt{2}}

\newcommand{\crn}{\nonumber \\}
\newcommand{\be}{\begin{equation}}
\newcommand{\ee}{\end{equation}}
\newcommand{\bea}{\begin{eqnarray}}
\newcommand{\eea}{\end{eqnarray}}
\newcommand{\ba}{\begin{eqnarray*}}
\newcommand{\ea}{\end{eqnarray*}}
\newcommand{\lbl}[1]{\label{eq:#1}}

\newcommand{\VVdA}{\langle VV\partial A \rangle}

\newcommand{\epo}{\;\:.}

\usepackage{graphicx}
\usepackage[figuresright]{rotating}

\title{
Non-renormalization of the full $\langle VVA \rangle$ correlator at
two--loop order
}

\author{F.~Jegerlehner~\address{
 Humboldt-Universit\"at zu Berlin, Institut f\"ur Physik,
       Newtonstrasse 15, D-12489 Berlin, Germany}
     and   O.~V.~Tarasov~\address{
     Deutsches Elektronen - Synchrotron, DESY,
       Platanenallee 6, D-15738 Zeuthen, Germany} }

\begin{document}
\onecolumn{
\renewcommand{\thefootnote}{\fnsymbol{footnote}}
\setlength{\baselineskip}{0.52cm}
\thispagestyle{empty}
\begin{flushright} \begin{tabular}{c}
HU-EP-06/05\\
DESY-06-022\\
SFB/CPP-06-04\\
January 2006 \end{tabular}
\end{flushright}

\setcounter{page}{0}

\mbox{}
\vspace*{\fill}
\begin{center}
{\Large\bf 
Non-renormalization of the full $\langle VVA \rangle$ correlator 
} \\
\vspace{3mm}
{\Large\bf at two--loop order} \\

\vspace{5em}
\large
F.~Jegerlehner$^a$ and 
O.~V.~Tarasov$^b$\footnote[1]{\noindent Talk given by O.T. at RADCOR 2005:
7th International Symposium On Radiative Corrections: Application Of
Quantum Field Theory To Phenomenology (RADCOR 2005), Shonan Village,
Japan, 2-7 Oct 2005. This work was supported by DFG
Sonderforschungsbereich Transregio 9-03 and in part by the European
Community's Human Potential Program under contract HPRN-CT-2002-00311
EURIDICE and the TARI Program under contract RII3-CT-2004-506078.}
\\[6mm]
\end{center}
\normalsize
$^a${\it   Humboldt-Universit\"at zu Berlin, Institut f\"ur Physik, 
Newtonstrasse 15, D-12489 Berlin, Germany}\\[4mm]
$^b${\it  Deutsches Elektronen-Synchrotron,
 DESY, Platanenallee 6, D--15738 Zeuthen, Germany}\\
\vspace*{\fill}}
\newpage

\begin{abstract}
\noindent
By explicit calculation of the two--loop QCD corrections we show that
for singlet axial and vector currents the full off--shell $\langle VVA
\rangle$ correlation function in the limit of massless fermions is
proportional to the one--loop result, when calculated in the
$\overline{\rm MS}$ scheme. By the same finite renormalization which
is needed to make the one--loop anomaly exact to all orders, we arrive
at the conclusion that two--loop corrections are absent altogether,
for the complete correlator not only its anomalous part. In accordance
with the one--loop nature of the $\langle VVA \rangle$ correlator, one
possible amplitude, which seems to be missing by accident at the
one--loop level, also does not show up at the two--loop level.
\end{abstract}

\maketitle

\section{INTRODUCTION}
\noindent
Recently, Vainshtein~\cite{Vainshtein03} found an important new
relation between form factors of the $\langle VVA \rangle$ correlator
matching to all orders in perturbation theory, in some kinematic
limit, a transversal amplitude to the anomalous longitudinal one,
which is known to be subject to the Adler-Bardeen non-renormalization
theorem~\cite{AdlerBardeen69}. Later Knecht et al.~\cite{KPPdR04} were
confirming this kind of non-renormalization theorem.  These recent
investigations came up in connection with problems in calculating the
leading hadronic effects in the electroweak two--loop contributions to
the muon anomalous magnetic moment
$a_\mu$~\cite{PPdR95,CKM95,KPPdR02,CMV03}.

The diagrams which yield the leading
corrections are those including a VVA triangular fermion--loop
($VVA\neq0$ while $VVV=0$ ) associated with a $Z$ boson exchange \\[-7mm]
\begin{center}
\includegraphics[height=2cm]{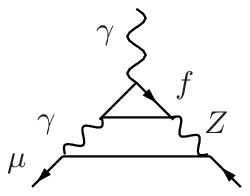}
\end{center}
\vspace*{-3mm}
and a fermion of flavor $f$ gives a potentially large contribution, up to UV singular terms which
will cancel~\cite{KKSS92},
\bea
a^{(4)\:\mathrm{EW}}_\mu([f])
\simeq ~~~~~~&& \\ && \hspace{-2.4cm}  \frac{\wz G_\mu m_\mu^2}{16 \pi^2}
\frac{\alpha}{\pi}\:2T_fN_{cf}Q_f^2\:\left[3 \ln \frac{M_Z^2}{m_{f'}^2} + C_f
\right] \nonumber
\eea
where $\alpha$ is the fine structure constant, $G_\mu$ the Fermi
constant, $T_{3f}$ the 3rd component of the weak isospin, $Q_f$ the
charge and $N_{cf}$ the color factor, 1 for leptons, 3 for quarks. The
mass $m_{f'}$ is $m_\mu$ if $m_f < m_\mu$ and $m_f$ if $m_f >
m_\mu$. $C_f$ denotes constant terms.  Since, as granted in the
Standard Model of elementary particles, anomaly cancellation by
lepton--quark duality $\sum_fN_{cf}Q_f^2T_{3f}=0$ is at work, only the
sums over complete lepton--quark families yield meaningful results
relevant to physics.  In any case the quark contributions have to be
taken into account, and treating them as free fermions the leading
large log $\sim \ln M_Z$ drops in sum over each family due to the
anomaly cancellation condition of the SM.

However, quarks cannot be treated perturbative and we expect
substantial strong interaction effects. A framework to investigate the
latter is to consider the general structure of the VVA three point function
\bea
\label{Greencalw}
&&{\cal W}_{\mu\nu\rho}(q_1,q_2) = i\int d^4x_1 d^4x_2
\,e^{i(q_1\cdot x_1 + q_2\cdot x_2)}
\,
\nonumber \\
&&
\times\,\langle\,0\,\vert\,\mbox{T}\{V_\mu(x_1)V_\nu(x_2)A_\rho(0)\}
\,\vert\,0\,\rangle
\eea 
of the flavor and color diagonal fermion currents 
\be V_{\mu} \,=\,{\overline\psi}\gamma_\mu\,\psi \quad ,
\quad A_{\mu} \,=\,
{\overline\psi}\gamma_\mu\gamma_5\,\psi 
\ee
where $\psi$ is a quark field.

To leading order the correlator
of interest is associated with the one--loop triangle diagram 
\begin{center}
\begin{picture}(60,40)(50,160)
\ArrowLine(65,165)(105,185)
\ArrowLine(105,185)(105,145)
\ArrowLine(105,145)(65,165)
\Photon(50,165)(65,165){2}{3}
\Photon(105,185)(120,185){2}{3}
\Photon(105,145)(120,145){2}{3}
\Vertex(65,165){2}
\Vertex(105,185){2}
\Vertex(105,145){2}
\Text(53,175)[]{$A_\rho$}
\Text(114,196)[]{$V_\mu$}
\Text(114,135)[]{$V_\nu$}
\Text(140,185)[]{$\leftarrow q_1 $}
\Text(140,145)[]{$\leftarrow q_2 $}
\end{picture}
\end{center}

\vspace{9mm}

\noindent
plus its crossed ($q_1,\mu \leftrightarrow q_2,\nu$) partner.

For the static low energy quantity $a_\mu=\ha
(g-2)_\mu=F_\mathrm{M}(0)$, given by the Pauli form factor at zero
momentum transfer, the VVA correlator is required in the limit
\bea \label{calwVL}{{\cal W}}_{\mu\nu\rho}(q_1=k+q,q_2=-k) &=&
 \\ && \hspace*{-4.5cm}
 -\,\frac{1}{8\pi^2}\,\bigg\{
w_L\left(q^2,0,q^2\right)\,q_\rho\, \veps_{\mu\nu\alpha\beta}\
q^\alpha k^\beta \nonumber\\
&& \hspace*{-3.6cm} +\,
w_T\left(q^2,0,q^2\right)\,t_{T\:\mu\nu\rho}
\bigg\}+O(k^2)\,,  \nonumber
\eea 
with {\footnotesize
$$t_{T\:\mu\nu\rho}=\left\{q^2 \veps_{\mu \nu \rho
\sigma}k^\sigma +q_\mu \veps_{\rho \nu \alpha \beta} q^\alpha k^\beta
-q_\rho \veps_{\mu \nu \alpha \beta} q^\alpha k^\beta \right\}\:.$$}
Indeed, in this kinematic region, the leading strong interaction
effects may be parametrized by two VVA amplitudes, a longitudinal
$w_L(Q^2)$ and a transversal $w_T(Q^2)$ one as functions of
$Q^2=-q^2$, which contribute as~\cite{KPPdR02,CMV03}
\bea
\Delta a_\mu^{(4)\:\mathrm{EW}}([f])_{\mathrm{VVA}}  &\simeq&
\frac{\wz G_\mu\:m_\mu^2}{16 \pi^2}\:\frac{\alpha}{\pi}\:\times~~~~~~~~~~~  \\
&& \hspace*{-3.3cm}
\int_{m_\mu^2}^{\Lambda^2}\:d Q^2\:\left(
w_L(Q^2)+ \frac{M_Z^2}{M_Z^2+Q^2}\:w_T(Q^2)\right)\;,\nonumber 
\eea 
where $\Lambda$ is a cutoff to be taken to $\infty$ at the end.
Vainshtein~\cite{Vainshtein03} has shown that in the chiral limit the
relation
\bea
\left. w_T(Q^2)_\mathrm{pQCD}\right|_{m=0}=\ha 
\left. w_L(Q^2)\right|_{m=0}\;,
\label{nonrentrans}
\eea
which was known to hold at one--loop~\cite{Rosenberg63}, is valid
actually to all orders of perturbative QCD.  Vainshtein's theorem
follows from the symmetry ($\rho,q \leftrightarrow \mu,q+k$) ($k \to
0$). Formally, discarding regularization problems, the asymptotic
symmetry derives from the fact that $\gamma_5$ may be moved from the
$A_\rho$ vertex to the $V_\mu$ vertex by anticommuting it an even
number of times~\cite{Vainshtein03}.  Thus for the quarks the
non-renormalization theorem valid beyond pQCD for the anomalous
amplitude $w_L$
\ba 
\left. w_L(Q^2)\right|_{m=0}=
\left. w^{1-\mathrm{loop}}_L(Q^2)\right|_{m=0} \!\!\! \!\!= \!\sum_q \:  \frac{4N_cT_qQ_q}{Q^2}
\ea
carries over to the perturbative part of the transversal amplitude.
Thus in the chiral limit the perturbative QPM result for $w_T$ is
exact. This may be somewhat puzzling, since in low energy effective
QCD, which encodes the non-perturbative strong interaction effects,
this kind of term seems to be absent. The term is recovered however by
taking into account all relevant terms in the operator product
expansion~\cite{CMV03,KPPdR04}.

In Vainshtein's kinematic limit, what matters is the derivative with
respect to $k$ taken at $k=0$. In this case actually the vertex
problem reduces to a propagator type problem. In the calculation
described below we have extended this to a genuine vertex type
statement at the two--loop level.  As the extensions of the
Adler-Bardeen non-renormalization theorem for the anomalous Ward
identity $\VVdA$ turn out to play an important role in new
phenomenological applications, we will study in the following such
possible generalizations by an explicit calculation of the leading QCD
corrections to the $\gamma \gamma Z$ triangle.\\

 The vector currents are strictly conserved $\partial_\mu
V^\mu=0$, while the axial vector current satisfies a PCAC relation
plus the anomaly $\partial_\mu A^\mu=2im_0 \bar{\psi}\gamma_5 \psi
+\frac{\alpha_0}{4\pi}\veps_{\mu \nu \rho \sigma}F^{\mu \nu}(x)F^{\rho
\sigma}(x)$.  
We will be mainly interested in the properties of strongly interacting
quark flavor currents in perturbative QCD.  Our notation closely
follows~\cite{KPPdR04}.

\noindent  The Ward identities restrict the general covariant 
decomposition of ${{\cal W}}_{\mu\nu\rho}(q_1,q_2)$ into invariant
functions to four terms
\bea \label{calw}-8\pi^2\,{{\cal W}}_{\mu\nu\rho}(q_1,q_2) &=&
 \nonumber \\ && \hspace*{-2.5cm}
~w_L\left(q_1^2,q_2^2,q_3^2\right)\,q_{3\rho}\, \veps_{\mu\nu\alpha\beta}\
q_1^\alpha q_2^\beta \nonumber\\
&& \hspace*{-2.5cm}+\,
w_T^{(+)}\left(q_1^2,q_2^2,q_3^2\right)\,t^{(+)}_{\mu\nu\rho}(q_1,q_2)
\nonumber \\ && \hspace*{-2.5cm}
+\,w_T^{(-)}
 \left(q_1^2,q_2^2,q_3^2\right)\,t^{(-)}_{\mu\nu\rho}(q_1,q_2)
\nonumber \\ && \hspace*{-2.5cm} 
 +\,
{\widetilde{w}}_T^{(-)}\left(q_1^2,q_2^2,q_3^2\right)\,{\widetilde{t}}^{(-)}_{\mu\nu\rho}(q_1,q_2)
\,,   \eea 
with $-q_3=q_1+q_2$ and transverse tensors given by 
\ba 
t^{(+)}_{\mu\nu\rho}(q_1,q_2)  =
q_{1\nu}\,\veps_{\mu\rho\alpha\beta}\ q_1^\alpha q_2^\beta \,&&
\hspace*{-7mm} -\,
q_{2\mu}\,\veps_{\nu\rho\alpha\beta}\ q_1^\alpha q_2^\beta \,
\nonumber \\ && \hspace*{-5.1cm}
-\, q_1 q_2\,\veps_{\mu\nu\rho\alpha}\ (q_1 - q_2)^\alpha
 +\ \frac{2\,q_1q_2}{q_3^2}\
\veps_{\mu\nu\alpha\beta}\ q_1^\alpha q_2^\beta q_{3 \rho}\ 
\nonumber  \\
t^{(-)}_{\mu\nu\rho}(q_1,q_2)  =\hspace{2.4cm}&&
\nonumber \\ &&\hspace*{-3.9cm}
\left[ (q_1 - q_2)_\rho \,+\, \frac{q_1^2 - q_2^2}{q_3^2}\,q_{3\rho} \right] \,\veps_{\mu\nu\alpha\beta}\ q_1^\alpha q_2^\beta
\nonumber\\
{\widetilde{t}}^{(-)}_{\mu\nu\rho}(q_1,q_2)  = 
q_{1\nu}\,\veps_{\mu\rho\alpha\beta} \ q_1^\alpha q_2^\beta \,&&
\nonumber \\ &&\hspace*{-3.2cm}
+\, q_{2\mu}\,\veps_{\nu\rho\alpha\beta}\ q_1^\alpha q_2^\beta
\,
+\, q_1 q_2\,\veps_{\mu\nu\rho\alpha}\ q_3 ^\alpha \,. \lbl{tensors}
\ea 
The longitudinal part is entirely fixed by the anomaly, 
\be
w_L\left(q_1^2,q_2^2,q_3^2\right)\,=\, - \frac{2N_c}{q_3^2}\,
\lbl{w_Lexpr} 
\ee 
which is exact to all orders of perturbation theory, the famous
Adler-Bardeen non-renormalization theorem.
The Vainshtein relation is obtained in the limit (\ref{calwVL})
upon identifying 
\ba
w_L(Q^2) &\equiv& w_L(q^2,0,q^2)\;,\;\; \\ w_T(Q^2) &\equiv&
w_T^{(+)}(q^2,0,q^2)+\widetilde{w}_T^{(-)}(q^2,0,q^2)\;, \nonumber
\ea 
with $Q^2=-q^2$.

\section{THE CALCULATION}
\noindent
Here we report on recent progress we made in extending
non-renormalization phenomenon at the two--loop
level~\cite{Jegerlehner:2005fs} .  For details and further references
we refer to the latter paper  in the following.  We perform the
calculation with conventional dimensional regularization in $d=4-2\ep$ 
dimensions and use a
linear covariant gauge with arbitrary gauge parameter throughout
the calculation.  Our procedure of treating $\gamma_5$ is similar to
the one used in~\cite{Jones:1982zf}. We write down all fermion loops
starting with the axial-vector vertex, and then perform Feynman
integrals and Dirac algebra without assuming any property of
$\gamma_5$ at all. In this way all diagrams will be expressed in terms
of traces of 10 combinations of $\gamma$ matrices. The prescription is
sufficient to enable us to arrive at amplitudes which have finite
limits as $d\rightarrow 4 $ in the corresponding covariant
decomposition.  After this the usual formulas
\ba
&&
{\rm Tr} [\gamma_5 \gamma_{\alpha}\gamma_{\beta}\gamma_{\mu}\gamma_{\nu}]
=4i\veps_{\alpha \beta \mu \nu}\;,\;\;
{\rm Tr} [\gamma_5 \gamma_{\alpha}\gamma_{\beta}] = 0
\ea
valid in $d=4$ dimensions were used. In our convention
$\veps_{0123}=+1$ and $(1-\gamma_5)/2$ projects to left--handed
fermion fields.




Tensor integrals were expressed in terms of integrals with different
shifts of the space-time dimension~\cite{Tarasov:1996br}.  All scalar
integrals could be reduced to 6 master integrals by using the Gr\"obner
basis technique proposed in~\cite{Tarasov:1998nx}.  The expressions
for the individual diagrams are sums over 21 terms which are
combinations of the 6 master integrals
\begin{eqnarray}
&&I_2^{(d)}(q_1^2) =\int\!\frac{ \widetilde{ d^d k_1}}{D_1D_3},
~~~~~~~~~~~~~\widetilde{d^d k_j} = \frac{d^d k_j}{ i \pi^{d/2}},
\nonumber \\
&&I_3^{(d)}(q_1^2,q_2^2,q_3^2) =\int  
\!\frac{\widetilde{ d^d k_1}}{D_1 D_3 D_4}
\nonumber \\
&&J_3^{(d)}(q_1^2) =\int\!\int 
\frac{\widetilde{d^d k_1} \widetilde{d^d k_2}}
{D_1 D_5 D_6},
\nonumber \\
&&R_1(q_1^2,q_2^2,q_3^2)=\int\!\int 
\frac{\widetilde{d^d k_1} \widetilde{d^d k_2} }
{D_1 D_5 D_6 D_7}
\nonumber \\
&&R_2(q_1^2,q_2^2,q_3^2)=\int\!\int 
\frac{\widetilde{d^d k_1} \widetilde{d^d k_2}}
{D_1^2 D_5  D_6 D_7},
\nonumber \\
&&P_5(q_1^2,q_2^2,q_3^2)=\int\!\int 
\frac{\widetilde{d^d k_1}\widetilde{d^d k_2}}
{D_1D_2 D_5 D_3 D_7},
\label{planarbasis}
\end{eqnarray}
multiplied by ratios of polynomials in $q_j^2$ and $d$.
Here $D_1=k_1^2$,  $D_2=k_2^2$, $D_3=(k_1-q_1)^2$,  $D_4=(k_1+q_2)^2$,
$D_5=(k_1-k_2)^2$, $D_6=(k_2-q_1)^2$ and $D_7=(k_2+q_2)^2$.
The integrals (\ref{planarbasis}) form a complete set of master
integrals needed for the calculation of massless vertex diagrams with
planar topology.
The planar integral with 6 denominators 
$$
P_6(q_1^2,q_2^2,q_3^2)=\int\!\int 
\frac{\widetilde{d^d k_1}\widetilde{d^d k_2}}
{D_1 D_3 D_4 D_5 D_6 D_7}
$$
can be reduced to integrals (\ref{planarbasis}) using
\begin{eqnarray} && \hspace*{-7mm}
q_3^2\ep P_6=
(1-2\ep) I_3(q_2^2,q_3^2,q_1^2) I_2^{(d)}(q_3^2)
\nonumber \\ &&\hspace*{-13mm}
 - R_2(q_1^2,q_2^2,q_3^2) 
 + R_2(q_1^2,q_3^2,q_2^2)
 + R_2(q_2^2,q_3^2,q_1^2)
\nonumber \\ && \hspace*{-7mm}
+ \ep ( P_5(q_1^2,q_3^2,q_2^2)
      + P_5(q_2^2,q_3^2,q_1^2)).
\nonumber 
\end{eqnarray}
$R_1$ satisfies the system of differential equations:
\ba
\{ x(1-x)\partial_x^2-y^2\partial_y^2
+[\gamma-(\alpha+\beta+1)x]\partial_x
\nonumber \\
-2xy\partial_x \partial_y
-(\alpha+\beta+1)y\partial_y-\alpha\beta \}R_1=0,
\nonumber \\
\{y(1-y)\partial_y^2 -x^2\partial_x^2
+[\gamma-(\alpha+\beta+1)y]\partial_y
\nonumber \\
 -2xy\partial_x\partial_y
-(\alpha+\beta+1)x\partial_x
-\alpha\beta \}R_1=0,
\ea
where $\beta=\gamma=2\alpha= 2\ep$ and
$$
  x=\frac{q_1^2}{q_3^2},
~~y=\frac{q_2^2}{q_3^2}, 
~~\partial_x= \frac{\partial}{\partial x},
~~\partial_y= \frac{\partial}{\partial y}.
$$
The general solution of this system may be written
in terms of Appell functions $F_4$ 
\bea
&&\hspace*{-13mm}(-q_3^2)^{2\ep}
R_1= 
AF_4(\ep,2\ep,2\ep,2\ep;x,y) \nonumber \\
&&\hspace*{-6mm}+BF_4(1-\ep,1,2-2\ep,2\ep;x,y)x^{1-2\ep} \nonumber  \\
&&\hspace*{-6mm}+CF_4(1-\ep,1,2\ep,2-2\ep;x,y)y^{1-2\ep} \nonumber  \\
&&\hspace*{-6mm}+DF_4(2-3\ep,2-2\ep,2-2\ep,2-2\ep;x,y),
\nonumber
\eea
where
\bea
&&\hspace*{-1cm}A =\frac{ \Gamma(2\ep) \Gamma^2(1-\ep) \Gamma(\ep) 
\Gamma(1-2\ep)}{(1-2\ep) \Gamma(2-3\ep)},
\nonumber \\
&&\hspace*{-1cm}B  = C=\frac{\Gamma(-1+2\ep) \Gamma^3(1-\ep)}
{(1-2\ep) \Gamma(2-3\ep)},
\nonumber \\
&&\hspace*{-1cm}D = \Gamma^2(1-\ep) \Gamma^2(-1+2\ep)x^{1-2\ep}y^{1-2\ep}.
\nonumber
\eea
The $F_4$ functions can be simplified yielding
\begin{eqnarray}
&&\hspace*{-1cm}(-q_3^2)^{2\varepsilon} R_1=
	    \frac{\Gamma(2\varepsilon) \Gamma^3\left(1-\varepsilon \right)}
{\varepsilon (1-2\varepsilon) \Gamma\left(2-3\varepsilon \right)}
\nonumber \\
&& \hspace*{-1cm}\times \left[   \frac{(Q_1+\lambda)}{2 y x^{2 \varepsilon-1}}
       F\left(\frac{Q_1+\lambda}{2y},\frac{Q_1+\lambda}{Q_1-\lambda}\right)
\right.
\nonumber \\
&&\hspace*{-1cm}\left.
+ \frac{(Q_2+\lambda)}{2xy^{2 \varepsilon-1}}
       F\left(\frac{Q_2+\lambda}{2x},
       \frac{Q_2+\lambda}{Q_2-\lambda} \right) \right]
\nonumber 
\\
&&\hspace*{-1cm}+\frac{ \pi \Gamma^2\left(\varepsilon-\frac12 \right)}
 {16^{1-\varepsilon} \sin(\pi \varepsilon)^2  }
\left[ 
 M~G \left( \frac{Q_3+\lambda}{2\lambda}\right)
\right.
\nonumber \\
&&\hspace*{-1cm} \left.
+\frac{2^{1+\ep} \cos(\pi \varepsilon)} 
     {(\lambda-Q_3)^{\varepsilon}}
 G\left( \frac{2\lambda}{\lambda-Q_3}\right)
\right]
\label{R1}
\end{eqnarray}
where $ ~~~\lambda= \sqrt{\Delta}~~~$ and
\bea
&&\hspace*{-10mm}\Delta=1+x^2+y^2-2xy-2x-2y, \nonumber \\
&&\hspace*{-12mm}Q_1=y\!+\!1\!-\!x,~Q_2=x\!+\!1\!-\!y,~Q_3=x\!+\!y\!-\!1, \nonumber 
\eea
\bea
&&\hspace*{-4mm}G(z)=\left._2F_1\right. \left(\ep,1-\ep,2-2 \ep,z \right),
\nonumber \\
&&\hspace*{-7mm}F(z,\omega)=F_1\left(1,1-\ep,1-\ep,1+\ep; z,\omega \right),
\nonumber 
\eea
\begin{eqnarray}
&&M = 
  \left( \frac{Q_3+\lambda}{-2 \lambda}\right)^{1-2\ep} 
   \frac{1}{\lambda^{\ep}} 
\nonumber \\
&& + \frac{x^{1-2 \ep} y^{1-2  \ep}}
    {\lambda^{1-  \ep}} 
    \left(\frac{Q_3+\lambda}{-2xy} \right)^{1-2 \ep}
\nonumber \\
&&
- \frac{(1-4 \cos(\pi \ep )^2)}{\lambda^{1-\ep}}
\left[ 
 \left(\frac{Q_3+\lambda}{2x}\right)^{1-2 \ep}\!\!\! x^{1-2 \ep} 
\right.
\nonumber \\
&&
\left.
+
 \left(\frac{Q_3+\lambda}{2y}\right)^{1-2 \ep} y^{1-2 \ep} 
\right].
\end{eqnarray}	    

The Gauss' hypergeometric function has an integral
representation 
\begin{equation}
G(z)=\frac{\Gamma(2-2\varepsilon)}{\Gamma^2(1-\varepsilon)}
\int_0^1\frac{du}{[u(1-u)(1-zu)]^{\varepsilon}}
\end{equation}
which is convenient for performing the $\varepsilon$ expansion.
For the expansion of $F(x,y)$  the relation
\begin{eqnarray}
&&\hspace*{-1cm}(1+\ep)(1-y)(1-x)(x-y) F(x,y)= \nonumber \\
&&\hspace*{-1cm}~~(x-y)[1 + (1-x)(1-y)\ep ] \phi(x,y)
\nonumber \\
&&\hspace*{-1cm}~~+x(x-y-x^2y+x^2)
\partial_x \phi(x,y)
\nonumber \\
&&\hspace*{-1cm}~~+y(x-y-y^2+xy^2) \partial_y \phi(x,y),
\end{eqnarray}
may be used to express this function in terms
of another $F_1$ function which is more suitable for $\ep$ expansion
\begin{eqnarray}
\phi(x,y) = F_1\left(1,-\ep,-\ep, 2+\ep;x,y\right)=
 \nonumber \\
(1+\ep) \int_0^1 du [(1-u)(1-xu)(1-yu)]^{\ep}.
\end{eqnarray}
An expression for $R_2$ may be obtained by differentiating the one
given for $R_1$.  The hypergeometric representation for $P_5$ is
obtained by solving the first order difference equation with respect
to $d$. Details of these calculations will be given in a separate
publication.  Series expansion in $\varepsilon$ for various master
integrals to the order needed in our calculations was given
in~\cite{Usyukina:1994iw,Usyukina:1994eg}. Recently, further terms of
the $\varepsilon$ expansion for these master integrals were calculated
in~\cite{Birthwright:2004kk}. 

The sum of all diagrams turns out to be
gauge parameter independent. In the Feynman gauge at $q_3=0$ and for 
arbitrary $d$, the results of our calculation are in agreement with 
the ones presented in~\cite{Jones:1982zf} diagram by diagram.

By applying the prescription outlined above for the evaluation 
of individual diagrams, and using Schouten's identities, we are able to
reshuffle terms to match  the tensor structures introduced in (\ref{calw}),
 exhibiting manifestly the vector current conservation.

\section{RESULTS AND DISCUSSION}

Including one-- and two--loop contributions, we may represent the 
form-factors in the form
\bea
&& w_T^{(\pm)}(q_1^2,q_2^2,q_3^2) =
 n_f\:N_c\: w_{1,T}^{(\pm)}(q_1^2,q_2^2,q_3^2)~
\nonumber \\ 
 &&+~ a~ n_f\:N_c\: C_2(R)\:  w_{2,T}^{(\pm)}(q_1^2,q_2^2,q_3^2)
\crn 
&&w_L(q_1^2,q_2^2,q_3^2) =   
 n_f\:N_c\: w_{1,L}(q_1^2,q_2^2,q_3^2)~
\nonumber \\
&&
 +~ a~ n_f\:N_c\: C_2(R)\:  w_{2,L}(q_1^2,q_2^2,q_3^2)
\label{formfactor}
\eea
where $ a = \alpha_s/(4\pi) $ 
includes the QCD coupling $\alpha_s$,
$n_f$ is the number of flavors and $N_c$ the number of colors.
The quarks are in the fundamental representation $R$ and the
corresponding group theory factor is
given by
\be
C_2(R) I = R^{a}R^{a}~~,~~~C_2(R)=4/3 ~~\mathrm{for \ QCD} \epo
\ee

We have been working in the $\overline{\rm MS}$ renormalization
scheme. The singlet axial current $J_{\rho}^5=A_\rho$ is non-trivially
renormalized because of the axial anomaly. It is
known~\cite{Trueman:1979en} that in addition to the standard
ultraviolet renormalization constant $Z_{\overline{\rm MS}}$ which
reads $Z_{\overline{\rm MS}}=1$ in our case (as $Z_{\overline{\rm
MS}}-1=O(a^2)$), one has to apply a finite renormalization constant
$Z_5$ such that renormalized and bare currents are related as:
\begin{equation}
(J_{\rho}^5 )_r = Z_5 Z_{\overline{\rm MS}}\: (J_{\rho}^5 )_0.
\end{equation}
The counterterms coming from the wave function renormalization of
quarks and ultraviolet renormalization of the axial and vector
currents cancel.  The finite renormalization constant is known at the
three-loop level~\cite{Larin:1993tq}.  For our calculations we must take
\begin{equation}
Z_5=1 - 4C_2(R)\:a \epo
\end{equation}

The result of the two--loop calculation after adding all diagrams is surprisingly simple and,
normalized according to (\ref{formfactor}), is given by 
\begin{eqnarray}
&&q_3^2 w_{2,L}(q_1^2,q_2^2,q_3^2) = -8,
\nonumber
\\
&& {\widetilde{w}}_{2,T}^{(-)}(q_1^2,q_2^2,q_3^2) = -  w_{2,T}^{(-)}(q_1^2,q_2^2,q_3^2),
\\
&&q_3^2 \Delta^2  w^{(-)}_{2,T}(q_1^2,q_2^2,q_3^2)
= 8 (x-y)\Delta
\nonumber \\
&&~~+8(x-y)(6xy + \Delta)\Phi^{(1)}(x,y)
\nonumber
\\
&&~~-4 [18 x y + 6 x^2-6 x + (1+x+y)\Delta)] L_x
\nonumber
\\
&&~~+ 4[18 x y + 6 y^2-6 y + (1+x+y)\Delta)] L_y
\nonumber
\\
\nonumber 
\\
&&q_3^2 \Delta^2  w^{(+)}_{2,T}(q_1^2,q_2^2,q_3^2)=
+8\Delta
\nonumber 
\\
&&
~~+8[6xy + (x+y)\Delta]\Phi^{(1)}(x,y) 
\nonumber \\
&&~~-4 [6 x+ \Delta ](x-y-1) L_x
\nonumber \\
&&~~+4 [6 y+ \Delta] (x-y+1) L_y
\end{eqnarray}
with  $L_x = \ln x,~~~~L_y = \ln y $.
Expression for $\Phi^{(1)}$ may be found  in~\cite{Usyukina:1992jd}:
\bea
\label{Phi1}
\Phi^{(1)} (x,y) = \frac{1}{\lambda} \left\{ \frac{}{}
2 \left( \Li{2}{-\rho x} + \Li{2}{-\rho y} \right)
\right.
\nonumber \\
\left.
+ \ln\frac{y}{x} \ln{\frac{1+\rho y}{1+\rho x}}
+ \ln(\rho x) \ln(\rho y) + \frac{\pi^2}{3}
\right\} ,
\eea
where
$$
\label{lambda}
\rho(x,y) \equiv 2 / (1-x-y+\lambda), 
$$
The comparison with the results of the one-loop calculation reveals that
\bea
{{\cal W}}_{\mu\nu\rho}(q_1,q_2)\left|_{two-loop}\right.=
\nonumber \\
4C_2(R) a {{\cal W}}_{\mu\nu\rho}(q_1,q_2)\left|_{one-loop}\right.
\eea

Multiplying the sum of one- and two-loop terms by the finite factor 
$Z_5$ we arrive at
\be
{{\cal W}}_{\mu\nu\rho}(q_1,q_2)
 = {{\cal W}}_{\mu\nu\rho}(q_1,q_2)\left|_{one-loop}\right.
\ee
This is the non-renormalization theorem for the full off shell
correlator at two--loops. While our calculation confirms the relations
derived in~\cite{KPPdR04} and the non-renormalization
theorem (\ref{nonrentrans}) found in~\cite{Vainshtein03,CMV03}, these findings are not
sufficient to explain our result valid for generic momenta.

Taking into account the rather non-trivial momentum dependence of the
form-factors it is very tempting to suggest that it could hold to all
orders of perturbation theory because of the topological nature of the
anomaly, for example. 

We would like to stress that the surprising relation could be
discovered only by keeping the general non-trivial momentum
dependence.  The anomalous three point correlator exhibits an
unusually simple structure, while contributions from individual
diagrams are very unwieldy.  One can expect similar effects for other
anomalous correlators. Since at the order considered the QCD
calculation is essentially a QED calculation, it is highly non-trivial
whether this carries over to higher orders. For the $\VVdA$ anomalous
correlator a large number of two--loop calculations have been
performed, mainly in QED
and we refer to the comprehensive review by Adler~\cite{Adler:2004ih}
and references therein.\\

In electroweak SM calculations one would a priori expect that
renormalizing parameters and fields would be sufficient for
renormalizing the SM. Our calculation shows that on top of the
standard renormalization, it is mandatory to renormalize the anomalous
currents $J^5_\rho$ by the finite renormalization factor $Z_5$ because
the lepton currents and the quark currents pick different
$Z$--factors and if they are not renormalized away the anomaly
cancellation and hence renormalizability obviously would get
spoiled. As pointed out by Adler and many others~\cite{Adler:2004ih}
the point is that there exists a renormalization scheme for which the
one--loop anomaly is exact. Only in this scheme anomaly cancellation
and thus renormalizability will carry over to higher orders in the
SM. Our result shows that due to the necessity of renormalizing away
possible higher order contributions from the anomaly also the
non-anomalous transversal contributions are affected. We have shown
that at least at two--loops the entire contribution gets renormalized
away in the zero mass limit.\\

\vspace*{1mm}
\noindent
{\bf Acknowledgments}\\

This work was supported by DFG Sonderforschungsbereich Transregio 9-03
and in part by the European Community's Human Potential Program under
contract HPRN-CT-2002-00311 EURIDICE and the TARI Program under contract 
RII3-CT-2004-506078. 
\noindent


\end{document}